\documentclass[
superscriptaddress,
reprint,
 amsmath,amssymb,
 aps,
prb,
]{revtex4-1}

\usepackage{graphicx,color}
\usepackage{dcolumn}
\usepackage{bm}

\begin{document}

\preprint{cond-mat}

\title{Weak antiferromagnetism of $J_\mathrm{eff}=1/2$ band in bilayer iridate Sr$_{3}$Ir$_{2}$O$_{7}$}
\author{S.\ Fujiyama}
\email{fujiyama@riken.jp} 
\affiliation{RIKEN, Advanced Science Institute, Wako 351-0198, Japan}
\author{K.\ Ohashi}
\affiliation{Department of Advanced Materials, University of Tokyo, Kashiwa 277-8561, Japan}
\author{H.\ Ohsumi}
\affiliation{RIKEN, SPring-8 Center, Sayo, Hyogo 679-5148, Japan}
\author{K.\ Sugimoto}
\affiliation{JASRI, Sayo, Hyogo 679-5198, Japan}
\author{T.\ Takayama}
\affiliation{Department of Advanced Materials, University of Tokyo, Kashiwa 277-8561, Japan}
\author{T.\ Komesu}
\affiliation{RIKEN, SPring-8 Center, Sayo, Hyogo 679-5148, Japan}
\author{M.\ Takata}
\affiliation{RIKEN, SPring-8 Center, Sayo, Hyogo 679-5148, Japan}
\author{T.\ Arima}
\affiliation{Department of Advanced Materials, University of Tokyo, Kashiwa 277-8561, Japan}
\affiliation{RIKEN, SPring-8 Center, Sayo, Hyogo 679-5148, Japan}
\author{H.\ Takagi}
\affiliation{Department of Physics, University of Tokyo, Hongo 113-0033, Japan}
\affiliation{RIKEN, Advanced Science Institute, Wako 351-0198, Japan}
\date{\today}

\begin{abstract}
The antiferromagnetic structure of Sr$_{3}$Ir$_{2}$O$_{7}$, the bilayer analogue of a spin-orbital Mott insulator Sr$_{2}$IrO$_{4}$, was revealed by resonant magnetic x-ray diffraction. Contrasting intensities of the magnetic diffraction at the Ir $L_\mathrm{III}$ and $L_\mathrm{II}$ edges show a $J_\mathrm{eff}=1/2$ character of the magnetic moment as is argued in Sr$_{2}$IrO$_{4}$. The magnitude of moment, however, was found to be smaller than that of Sr$_{2}$IrO$_{4}$ by a factor of 5-6, implying that Sr$_{3}$Ir$_{2}$O$_{7}$ is no longer a Mott insulator but a weak antiferromagnet. An evident change of the temperature dependence of the resistivity at $T_\mathrm{N}$, from almost temperature-independent resistivity to insulating, strongly suggests that the emergent weak magnetism controls the charge gap. The magnetic structure was found to be an out-of-plane collinear antiferromagnetic ordering in contrast to the inplane canted antiferromagnetism in Sr$_{2}$IrO$_{4}$, originating from the strong bilayer antiferromagnetic coupling.
\end{abstract}

\pacs{75.25.-j, 75.70.Tj, 71.30.+h}
\maketitle
Recently, novel interplay of spin-orbit coupling and electron correlations in heavy $5d$ transition metal oxides has been attracting considerable interest as a new paradigm of oxide physics. In Sr$_{2}$IrO$_{4}$, a layered Ir$^{4+}$ perovskite with five $d$-electrons in its $t_{2g}$ orbitals, such interplay is particularly pronounced. Sr$_{2}$IrO$_{4}$ is an insulator with a charge gap of the order of $\lesssim$ 0.5 eV and shows a canted antiferromagnetism below $T_\mathrm{N}\sim230$ K.~\cite{Crawford1994,Cao1998,Kim2008} The insulating behavior even well above $T_\mathrm{N}$ and the presence of a large localized magnetic moment of $\sim 0.5 \mu_{B}$, estimated from the large canted moment 0.075 $\mu_{B}$/Ir, indicate that Sr$_{2}$IrO$_{4}$ can be viewed as a Mott insulator. The Mottness of Sr$_{2}$IrO$_{4}$ is believed to be associated with a formation of a half-filled $J_\mathrm{eff}=1/2$ band, created by a very strong spin-orbit splitting as large as $\sim 0.7$ eV comparable to the width of the $t_{2g}$ band $\sim 1.5$ eV. The polarization dependence of x-ray absorption spectra and the contrasting resonances at $L_\mathrm{II}$ and $L_\mathrm{III}$ edges in the magnetic x-ray diffraction firmly evidence $J_\mathrm{eff}=1/2$ character of the magnetic moment.~\cite{Kim2008,Kim2009,Fujiyama2012} 

The Mottness of Sr$_{2}$IrO$_{4}$ characterized by the charge gap as small as $\lesssim 0.5$ eV is marginal. Like some other tetravalent transition metal ions such as Ti$^{4+}$, Mn$^{4+}$ and Ru$^{4+}$, Ir$^{4+}$ forms a series of perovskite-based structures called Ruddlesden-Popper series, Sr$_{n+1}$Ir$_{n}$O$_{3n+1}$, with stacking of $n$-IrO$_{{2}}$ layers as structural units. Reflecting the marginal Mottness in Sr$_{2}$IrO$_{4}$ ($n=1$), the system becomes more and more itinerant with increasing $n$. The three dimensional analogue of the series, SrIrO$_{3}$ ($n=\infty$) is known to show a metallic transport.~\cite{Cao2007} Recent study revealed that the ground state of SrIrO$_{3}$ is a semimetal close to a band insulator,~\cite{TakayamaUnp} where the lifting of band degeneracy by spin-orbit coupling plays a key role in producing the incomplete charge gap.~\cite{Carter2012} Very likely due to the presence of strong spin orbit coupling, the semimetal state is located almost right next to the Mott insulator. A question arises how the Mott insulator evolves into the semimetal in Sr$_{n+1}$Ir$_{n}$O$_{3n+1}$.       

\begin{figure}[htb]
\includegraphics*[width=8.5cm]{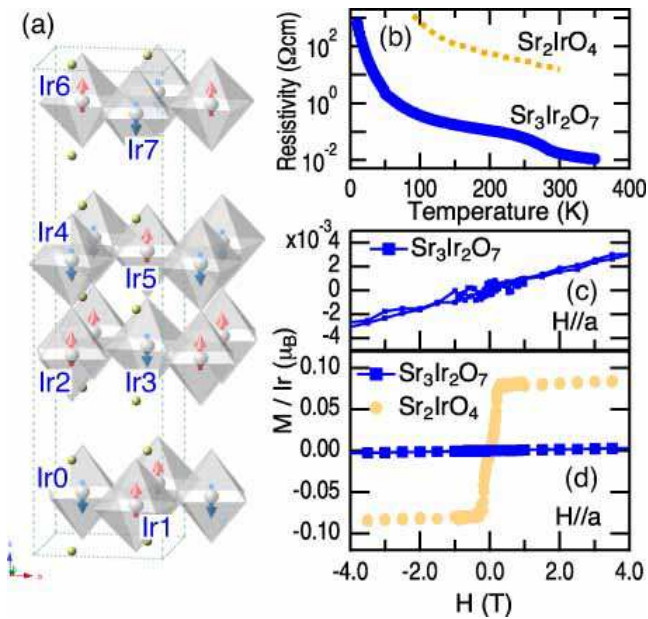}
\caption{(a) The crystal and magnetic structures of Sr$_{3}$Ir$_{2}$O$_{7}$. The dashed line denotes the unit cell. Note that the volume of the magnetic unit cell agrees to that of the lattice. We label eight iridium sites (grey circles, Ir${j}$, $j=0 \ldots 7$) for the following analysis of the magnetic structure. (b) Inplane resistivities of Sr$_{3}$Ir$_{2}$O$_{7}$ and Sr$_{2}$IrO$_{4}$. (c) and (d) Magnetizations of Sr$_{3}$Ir$_{2}$O$_{7}$ at 260 K and Sr$_{2}$IrO$_{4}$ at 5 K~\cite{MagCom}.}
\label{fig:structure}
\end{figure}

Sr$_{3}$Ir$_{2}$O$_{7}$, the $n=2$ member of Sr$_{n+1}$Ir$_{n}$O$_{3n+1}$, could be a key compound to pursue the electronic evolution from the spin-orbital Mott insulator to the semimetal. As shown in Fig.~\ref{fig:structure} (a), the crystal structure consists of stacking of bilayers of corner-shared IrO$_{6}$ octahedra along the $c$-axis. The octahedra are rotated around the $c$-axis and the inplane Ir-O-Ir bonds are hence bent. The directions of the rotations are opposite for the neighboring IrO$_{6}$ octahedra not only within each layer but also between the two layers. Sr$_{3}$Ir$_{2}$O$_{7}$ is reported to show a semiconducting temperature dependence of the resistivity but much more conducting than Sr$_{2}$IrO$_{4}$, implying close proximity to the critical border to the semimetal.~\cite{Cao2002,Subramanian1994,Nagai2007} A very weak ferromagnetic moment was observed in the temperature dependent magnetization below 280 K, suggesting the presence of magnetic ordering, but the magnetic structure has not been fully explored yet. By determining the magnetic structure of Sr$_{3}$Ir$_{2}$O$_{7}$, we should be able to address the following issues closely linked to the evolution of the electronic state from the Mott insulator to the semimetal.

Is the $J_\mathrm{eff}=1/2$ state robust against the change of the crystal structure and the itinerancy of the electrons? The ratios of the inplane and out-of-plane Ir-O bond lengths are 1.04 for Sr$_{2}$IrO$_{4}$ and 1.02 for Sr$_{3}$Ir$_{2}$O$_{7}$, implying less distorted IrO$_{6}$ octahedra.~\cite{Cao2002} Assuming a comparable spin-orbit coupling to Sr$_{2}$IrO$_{4}$, we may anticipate a formation of almost ideal $J_\mathrm{eff}=1/2$ state. However, it is highly nontrivial whether the increased itinerancy could destabilize the $J_\mathrm{eff}=1/2$ state, which should be examined experimentally. Even if the $J_\mathrm{eff}=1/2$ character holds for Sr$_{3}$Ir$_{2}$O$_{7}$, the estimate of the magnitude of the ordered moment will provide a key to examine the Mottness of this material.

The magnetic structure of this material should be pinned down. The rotational distortion of the IrO$_{6}$ octahedra, alternating in each plane and opposite between the upper and the lower layers, could give rise to a net ferromagnetic canting moment when antiferromagnetically coupled moments are lying in the plane through Dzyaloshinskii-Moriya (DM) interaction as is argued in Sr$_{2}$IrO$_{4}$. On the other hand, the intrabilayer coupling, $J_{\perp}$, that is absent in Sr$_{2}$IrO$_{4}$ and perhaps antiferromagnetic that is comparable or even larger than that within the plane could conflict with the inplane canted moments. The reported weak ferromagnetic moment of Sr$_{3}$Ir$_{2}$O$_{7}$ was in fact only 1\% of that of Sr$_{2}$IrO$_{4}$ ($<10^{-3}\mu_{B}$/Ir) and showed very peculiar temperature dependence.~\cite{Cao2002} We suspect that this could be originating from an extrinsic moment produced by defects.

In this communication, we report the magnetic structure of the bilayer perovskite, Sr$_{3}$Ir$_{2}$O$_{7}$, determined by resonant magnetic x-ray diffraction. The magnetic structure was found to be collinear along the $c$-axis with antiferromagnetic intrabilayer coupling, in marked contrast to the inplane canted antiferromagnetism in Sr$_{2}$IrO$_{4}$. We found that the ordered moment below $T_\mathrm{N}=280$ K has a $J_\mathrm{eff}=1/2$ character but that the magnitude is substantially reduced to 1/5-1/6 of that of Sr$_{2}$IrO$_{4}$, which points out that Sr$_{3}$Ir$_{2}$O$_{7}$ is a $J_\mathrm{eff}=1/2$ band magnetic semiconductor rather than a $J_\mathrm{eff}=1/2$ Mott insulator.

A single crystal of Sr$_{3}$Ir$_{2}$O$_{7}$ with a dimension of 2 mm$\times $2 mm $\times$0.5 mm was synthesized by a flux method. The resistivity measurement was conducted by a conventional four probe technique. The magnetization measurement was performed using a commercial SQUID magnetometer. A single crystal x-ray diffraction measurement for the determination of the lattice structure was performed using a cylindrical IP (imaging plate) on BL02B1 at SPring-8.~\cite{Sugimoto2010} The wavelength of the incident x-ray was set to 0.354 \AA. Resonant magnetic diffraction measurement using a multi-circle diffractometer was performed on BL19LXU of SPring-8.~\cite{Yabashi2001} The wavelength was set to 1.106 \AA\  corresponding to the $L_\mathrm{III}$ edge ($2p_{3/2}\rightarrow 5d$) of iridium for the magnetic structure analysis. The $b$-axis of the crystal was nearly kept parallel to the scattering plane.

The analysis of the crystal structure indicates an orthorhombic structure with the space group $Bbeb$ and the lattice constants of 5.5108, 5.512, 20.8832 \AA, that has lower symmetry than $I\bar{4}/mmm$ that is reported at very early stage,\cite{Subramanian1994} but agrees to more recent analysis.~\cite{Cao2002} The result of structural refinement is shown in Table~\ref{tab:atom}. We found that the presence of the two domains associated with orthorhombic distortion and that the ratio of the volumes of the two domains was roughly 0.7:0.3. In the magnetization of the crystal measured, a very weak ferromagnetism was observed below 280K only when the inplane magnetic field was applied. The magnitude of moment, however, is almost 1\% of that of Sr$_{2}$IrO$_{4}$ as shown in Fig.~\ref{fig:structure}(c) and (d), in agreement with the previous report. On the other hand, the reported peculiar reduction of the susceptibility below 50 K was not reproduced.~\cite{Cao2002} Additional weak ferromagnetic signal was observed below 230K, which we attribute to the intergrowth of Sr$_{2}$IrO$_{4}$ layer.~\cite{Boseggia2012} From the magnetization measurement, we estimate that the intergrowth is 1 \% of the total volume.
The resistivity of the crystal in Fig.~\ref{fig:structure}(b) showed a semiconducting temperature dependence. An evident anomaly at the magnetic transition temperature is observed suggestive of a close interrelation between the magnetism and the motion of electrons. Above the magnetic transition temperature, $T_\mathrm{N}\sim 280$ K, the magnitude of the resistivity is as low as $\rho\sim 10$ m$\Omega$cm and weakly temperature dependent. This suggests that the system is a poor metal or a very narrow gap semiconductor at high temperatures.

 \begin{table}
 \caption{\label{tab:BL2}Atomic positions of Sr$_{3}$Ir$_{2}$O$_{7}$ determined by synchrotron radiation x-ray diffraction at 295 K.}
 \begin{ruledtabular}
 \begin{tabular}{cccc}
atom & $x$ & $y$ & $z$ \\ \hline
 Ir & 0.2500 & 0.7500 & 0.0974 \\
 Sr1 & 0.2500 & 0.2500 & 0.0000 \\
 Sr2 & 0.2500 & 0.2500 & 0.2500 \\
 O1 & 0.2500 & 0.7500 & 0.0000 \\
 O2 & 0.2500 & 0.7500 & 0.1943 \\
 O3 & 0.5487 & 0.5485 & 0.0962 \\
  \end{tabular}
  \label{tab:atom}
 \end{ruledtabular}
 \end{table}

\begin{figure}[htb]
\includegraphics*[width=8cm]{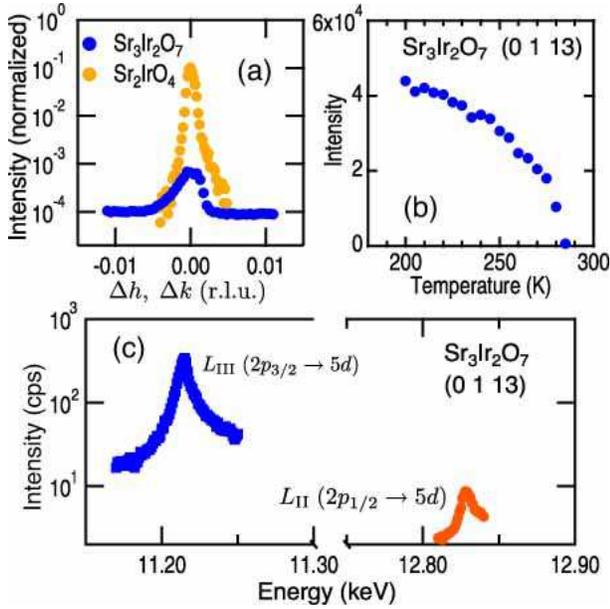}
\caption{(a) The resonant magnetic diffraction normalized by the charge scattering of Sr$_{2}$IrO$_{4}$ ($I_{m}$(1 0 22)/$I_{c}$(0 0 24)) and Sr$_{3}$Ir$_{2}$O$_{7}$ ($I_{m}$(0 1 13)/$I_{c}$(0 0 20)). (b) Temperature evolution of the magnetic diffraction (0 1 13). (c) Strong enhancement of the diffraction at the $L_\mathrm{III}$ edge ($c\hbar/\lambda=11.2 $ keV) is observed while only small diffraction is observed at the $L_\mathrm{II}$ edge ($c\hbar/\lambda=12.8$ keV).}
\label{fig:Jonehalf}
\end{figure}

How about magnetism? We found the (0 1 13) reflection in the IP at 30 K when the incident wavelength of x-ray was tuned at the $L_\mathrm{III}$ edge of iridium, that is prohibited in a $B$-base centered lattice and apparently shows an inplane antiferromagnetism. To elucidate the magnetic correlation along the $c$-axis, particularly the unknown intrabilayer coupling, we performed a resonant magnetic x-ray diffraction experiment by a four-circle diffractometer for (0 1 $l$) reflections. The magnetic diffraction evolves below 280 K as shown in Fig.~\ref{fig:Jonehalf} (b), which evidences that the anomaly in the magnetization originates from the antiferromagnetic long range ordering. Upon cooling, the magnetic diffraction evolves monotonously, excluding a possibility of the second magnetic transition at $\sim 230$ K. The magnetic diffraction that is normalized by the intensity of the charge diffraction ($I_{m}(0\, 1\, 13)/I_{c}(0\, 0 \, 20)$) is reduced to nearly 1/150 to that of a spin-orbital Mott insulator Sr$_2$IrO$_4$ ($I_{m}(1 \, 0 \, 22)/I_{c}(0 \, 0 \, 24)$) as shown in Fig.~\ref{fig:Jonehalf} (a) plausibly originating from the itinerant character of electrons. It is to be noted that the structure factors of the charge diffractions, $F_{0 \, 0 \, 20}$ of Sr$_{3}$Ir$_{2}$O$_{7}$ and $F_{0 \, 0 \, 24}$ of Sr$_{2}$IrO$_{4}$, have nearly the same magnitudes. Despite the contrasting difference in the diffraction intensities, the ordered moment keeps a $J_\mathrm{eff}=1/2$ character like Sr$_2$IrO$_4$ evidenced by the strong enhancement at the $L_\mathrm{III}$ edge as shown in Fig.~\ref{fig:Jonehalf} (c). 

\begin{figure}[htb]
\includegraphics*[width=8cm]{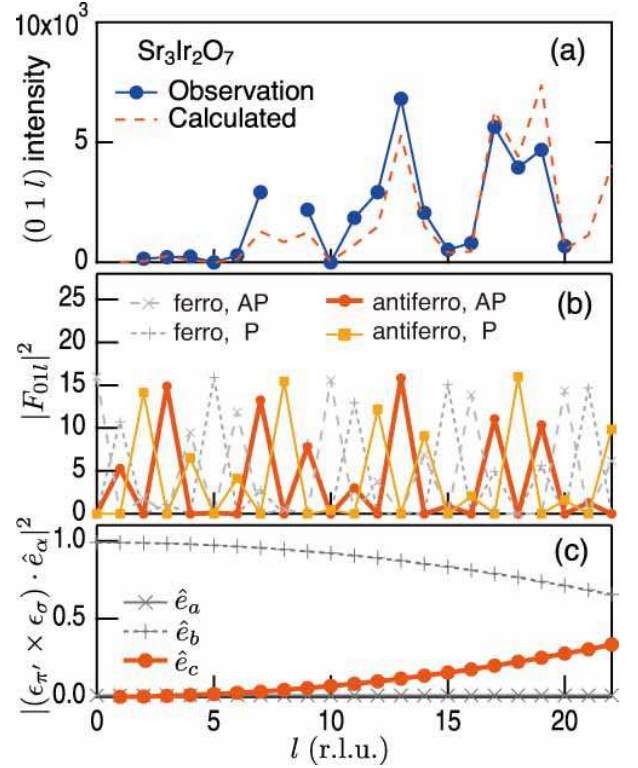}
\caption{(a) Intensities of the resonant magnetic x-ray diffraction at (0 1 $l$) (blue circle). We encountered a technical difficulty in measuring (0 1 8) reflection by the geometry of the diffractometer, and this is missing. Orange dashed line is the calculated intensity for the antiferromagnetic intrabilayer coupling with moment parallel to the $c$-axis assuming a mixture of 90 $^{\circ}$ rotated domains by a factor of 0.7 to 0.3. See the detail in the text.
(b) Calculated $|F_{01l}|^{2}$ of four possible magnetic structures with intrabilayer magnetic couplings ($f_{4}=\pm f_{2}$) and interbilayer antiparallel (AP) and parallel (P) relations ($f_{0}=\mp f_{2}$). (c) The $|(\bm{\epsilon_{\pi'}\times \bm{\epsilon_{\sigma}}})\cdot \bm{\hat{e}_{\alpha}}|^{2}$ for $\bm{\hat{e}_{\alpha}}\parallel a,b $ and $c$.}
\label{fig:Factor}
\end{figure}

We show in Fig.~\ref{fig:Factor} (a) the $l$ dependence of the magnetic diffraction.  
Under the resonance condition at the $L$ edge, the diffraction intensity is described as, 
\begin{equation*}
I (hkl)\propto |(\bm{\epsilon'}\times \bm{\epsilon})\cdot \bm{m}|^{2}|F_{hkl}|^{2}= |(\bm{\epsilon_{\pi'}}\times \bm{\epsilon_{\sigma}})\cdot \bm{m}|^{2}|F_{hkl}|^{2}, 
\end{equation*}
where $\bm{\epsilon (\epsilon')} $ is the polarization vector of the incident (scattered) x-ray, $\bm{m}=m_{a}\bm{\hat{e}}_{a}+m_{b}\bm{\hat{e}}_{b}+m_{c}\bm{\hat{e}}_{c}$ is the ordered moment and $F_{hkl}$ denotes the structure factor.

A continuous $l$ scan along (0 1 $l$) detects no magnetic superspot, which shows that the magnetic periodicity along the $c$-direction agrees to the $c$ of the unit cell containing four IrO$_{6}$ layers. When we denote the direction (sign) of the moment at Ir${j}$ site as $f_{j} (j=0\ldots7)$ (Fig.~\ref{fig:structure} (a)) under a condition of inplane antiferromagnetism for each IrO$_{2}$ plane evidenced by the IP measurement as, $f_{1 \, (3,5,7)}=-f_{0 \, (2,4,6)}$, the squared structure factor at (0 1 $l$) is calculated as,
\begin{eqnarray*}
|F_{01l}|^{2}&=&4[\sin^{2}\theta( f_{0} - f_{6}+(-1)^{l}f_{2}-(-1)^{l}f_{4})^{2}\\&+&\cos^{2}\theta(- f_{0} - f_{6}+(-1)^{l}f_{2}+(-1)^{l}f_{4})^{2}],
\end{eqnarray*}
using $\theta=2\pi\cdot 0.0974\cdot l$.

The possible magnetic structures are limited to 4 cases depending on the intrabilayer magnetic couplings ($J_{\perp}$, $f_{4}=\pm f_{2}$ for ferromagnetic (antiferromagnetic)) and the interbilayer relations ($f_{0}=\mp f_{2}$ for antiparallel (parallel)), of which $|F_{01l}|^{2}$ are plotted in Fig.~\ref{fig:Factor} (b). The sign of $J_{\perp}$ determines the phase of the overall $l$ dependence with a periodicity of $1/(2\cdot 0.0974)=5.13$, and the plot in Fig.~\ref{fig:Factor} (b) apparently shows an antiferromagnetic intrabilayer coupling.

The smearing-out of the even-odd alternating $l$-dependence of the observed diffraction is originating from the presence of two structural domains in the crystal. Using the ratio of the domains of 0.7 to 0.3 deduced by the crystal structure analysis, the observed intensity is expected as $I(01l)\propto0.7|(\bm{\epsilon_{\pi'}}\times\bm{\epsilon_{\sigma}})\cdot \bm{\hat{e}}_{\alpha}|^{2}|F_{01l}|^{2}+0.3|(\bm{\epsilon_{\pi'}}\times\bm{\epsilon_{\sigma}})\cdot \bm{\hat{e}}_{\beta}|^{2}|F_{10l}|^{2}$, where $\bm{\hat{e}}_{\beta}$ should be 90$^{\circ}$ rotated to $\bm{\hat{e}}_{\alpha}$ around the $c$-axis.~\cite{interCom} The calculated $|(\bm{\epsilon_{\pi'}}\times\bm{\epsilon_{\sigma}})\cdot \bm{\hat{e}}_{\alpha}|^{2}$ for $\alpha=a,b$ and $c$ show contrasting  $l$ dependences as well as the magnitudes as shown in Fig.~\ref{fig:Factor} (c). Under the condition of the sample orientation with $b$-axis parallel to the scattering plane, the $m_{a}\bm{\hat{e}}_{a}$ gives negligible contribution to the diffraction because of $\bm{\epsilon_{\sigma}} \parallel \bm{\hat{e}}_{a}$. The only choice of the unit vectors to smear out the even-odd alternation is $\bm{\hat{e}}_{\alpha}=\bm{\hat{e}}_{\beta}=\bm{\hat{e}}_{c}$, therefore, we conclude that the ordered moments are directed along the $c$-axis.~\cite{MomentCom} We plot in Fig.~\ref{fig:Factor} (a) the calculated $l$ dependence of the diffraction considering the structural domains and $c$-directed moments with antiferromagnetic $J_{\perp}$, which shows a good agreement with the observed magnetic diffraction. The determined magnetic structure as shown in Fig.~\ref{fig:structure} (a) well resolves the absence of weak ferromagnetic moment and the negligible magnetization below $T_\mathrm{N}$ ($<$ 0.001 $\mu_{B}$ at 260 K), because the moments lie parallel to the DM vectors. 

The most plausible source to realize the collinear magnetic structure along the $c$-axis is the antiferromagnetic $J_{\perp}$. In contrast to the inplane Ir-O-Ir bond that is bent with 11$^{\circ}$, the straight bond along the out-of-plane direction connecting two layers with nearly the same distance expects a considerable $J_{\perp}$ with nearly the same or even larger than the inplane antiferromagnetic exchange. We consider that this large $J_{\perp}$ can conflict with the DM interaction favoring inplane canting of the moments, and as a consequence to avoid the competition, the moments are directed parallel to the DM vectors.

The observed magnetic diffraction normalized by the charge diffraction is nearly 150 times smaller than that of Sr$_{2}$IrO$_{4}$. Even considering the reduction of the diffraction intensity originating from the difference of the directions of the moments by a factor of $[|(\bm{\epsilon_{\pi'}}\times\bm{\epsilon_{\sigma}})\cdot \bm{\hat{e}}_{a}|^{2}$ at (1 0 22)]/[$|(\bm{\epsilon_{\pi'}}\times\bm{\epsilon_{\sigma}})\cdot \bm{\hat{e}}_{c}|^{2}$ at (0 1 13)] $\sim 5$, a strong reduction of the squared moment $|m_{c}$ (Sr$_{3}$Ir$_{2}$O$_{7}$) /$m_{a}$ (Sr$_{2}$IrO$_{4}$)$|^{2}\sim 1/30$ is suggested. While the magnitudes of the ordered moments of different samples cannot precisely be determined solely by resonant x-ray diffraction, the moment of Sr$_{3}$Ir$_{2}$O$_{7}$ is considerably reduced to 1/5 - 1/6 of that of Sr$_{2}$IrO$_{4}$, i.e. $\sim 0.1$ $\mu_{B}$/Ir. The reduced magnetic moment clearly shows that the magnetism of Sr$_{3}$Ir$_{2}$O$_{7}$ should be better described by a band magnetism of $J_\mathrm{eff}=1/2$ electrons rather than a Mott insulator. The anomaly in the resistivity is very clearly observed at $T_\mathrm{N}$, implying the band magnetism is controlling the poor-metal insulator transition.  One may argue the possibility of the formation of a Slator insulator or an SDW state below $T_\mathrm{N}$.~\cite{Arita2012} We, however, note that the magnetic structure revealed has the same unit cell as the unit cell of the lattice and a band folding associated with the antiferromagnetic ordering would not be anticipated. In this respect, neither Slator nor SDW scenario in their most naive form can be applied here and much more elaborated picture should be invoked. It is interesting to note here again the contrast between Sr$_{2}$IrO$_{4}$ and Sr$_{3}$Ir$_{2}$O$_{7}$. Only a minute change in the crystal structure alters the system from a strongly localized magnet to a weakly localized band magnet, which we believe represents a extremely delicate interplay of Coulomb $U$, kinetic energy $t$, spin-orbit coupling and lattice.

In conclusion, we disclosed the antiferromagnetic structure of the bilayer perovskite iridium oxide, Sr$_{3}$Ir$_{2}$O$_{7}$, by resonant magnetic x-ray diffraction. Despite much more itinerant character of electrons than the spin-orbital Mott insulator, Sr$_{2}$IrO$_{4}$, $J_\mathrm{eff}=1/2$ electronic state holds for this material. The ordered magnetic moment is directed along the $c$-axis, which explains the absence of the remnant ferromagnetic moment, plausibly originating from the competition of the intrabilayer antiferromagnetic exchange and the Dzyaloshinskii-Moriya interaction. Apart from the resemblance of the antiferromagnetism and $T_\mathrm{N}$ to Sr$_{2}$IrO$_{4}$, the ordered moment is strongly reduced down to nearly 1/5 to that of Sr$_{2}$IrO$_{4}$, showing that the electronic state is no longer a Mott insulator but a band magnet. A considerable interplay between the magnetism and the transport is demonstrated, where the emergent moment appears to control the transition from a poor metal (narrow gap semiconductor) to an insulator. In the course of this study, we noticed a result of magnetic x-ray diffraction of this sample.~\cite{Kim2012} The magnetic structure proposed is consistent with the current study.

\begin{acknowledgments}
We are grateful to W.\ Koshibae, D.\ Hashizume, H.Y.\ Kee, G.\ Khaliullin, and G.\ Jackeli for fruitful discussions. The synchrotron radiation experiments were performed at BL19LXU in SPring-8 with the approval of RIKEN (20080047). This work was supported by Grant-in-Aid for Scientific Research on Priority Areas (19052007, 17071001) from MEXT.
\end{acknowledgments}


\begin{thebibliography}{}
\bibitem{MagCom}Since our single crystal of Sr$_{3}$Ir$_{2}$O$_{7}$ contains the intergrowth of Sr$_{2}$IrO$_{4}$, we show the magnetization data at 260 K, 20 K below $T_\mathrm{N}$.\ We consider nearly half of the saturated magnetization develops at 260 K.
\bibitem{MomentCom} Mathematically, there remains a possibility of $m_{a}=m_{b}$ corresponding to the direction of moments to $\bm{\hat{e}}_{a}+\bm{\hat{e}}_{b}$, however, this magnetic structure could not be realized in an orthorhombic lattice. In addition, inplane moments meet a significant conflict to the magnetization as discussed below.
\end{thebibliography}
\end{document}